\documentclass[aps,prl,twocolumn,groupedaddress]{revtex4}

\usepackage{amsmath}
\usepackage{amssymb}
\usepackage{graphicx}
\usepackage{bm}

\begin{document}

\title{Synchronous imaging for rapid visualization of complex vibration profiles in electromechanical microresonators}

\author{Y. Linzon}\email{yoli@braude.ac.il; yoav.linzon@cornell.edu.}
\affiliation{Department of Physics and Optical Engineering,
Ort Braude College, PO Box 78, Karmiel 21982, Israel}

\author{D. J. Joe}
\affiliation{School of Applied and Engineering Physics,
Cornell University, Ithaca, New York 14853, USA}

\author{S. Krylov}
\affiliation{School of Mechanical Engineering, Faculty of
Engineering, Tel Aviv University, Tel Aviv 69978, Israel}

\author{B. Ilic}
\affiliation{School of Applied and Engineering Physics,
Cornell University, Ithaca, New York 14853, USA}

\author{J. Topolancik}
\affiliation{School of Applied and Engineering Physics,
Cornell University, Ithaca, New York 14853, USA}

\author{J. M. Parpia}
\affiliation{School of Applied and Engineering Physics,
Cornell University, Ithaca, New York 14853, USA}

\author{H. G. Craighead}
\affiliation{School of Applied and Engineering Physics,
Cornell University, Ithaca, New York 14853, USA}

\date{\today}

\begin{abstract}

Synchronous imaging is used in dynamic space-domain vibration
profile studies of capacitively driven, thin n+ doped
poly-silicon microbridges oscillating at rf frequencies. Fast
and high-resolution actuation profile measurements of
micromachined resonators are useful when significant device
nonlinearities are present. For example, bridges under
compressive stress near the critical Euler value often reveal
complex dynamics stemming from a state close to the onset of
buckling. This leads to enhanced sensitivity of the vibration
modes to external conditions, such as pressure, temperatures,
and chemical composition, the global behavior of which is
conveniently evaluated using synchronous imaging combined
with spectral measurements. We performed an experimental
study of the effects of high drive amplitude and ambient
pressure on the resonant vibration profiles in
electrically-driven microbridges near critical buckling.
Numerical analysis of electrostatically driven post-buckled
microbridges supports the richness of complex vibration
dynamics that are possible in such micro-electromechanical
devices.

\end{abstract}


\maketitle

\section{Introduction}

Suspended resonant nano- and micro-electro-mechanical systems
(NEMS and MEMS) find use in versatile applications, such as
ultra-sensitive mass detectors, rf filters, and switching
devices \cite{Craig,Roukes1}. As device miniaturization
advances, optimization of the overall characteristics in
high-frequency MEMS/NEMS resonators becomes increasingly
complex and linked with various mechanical, electrical,
thermal and optical parameters of the system and its
environment. This compounds their seemingly superior
sensitivity to environmental conditions, such as the
pressure, temperature and chemical composition of the
surrounding gas.

In the characterization of NEMS and MEMS under periodic
electrical actuation, vibration profile (VP) measurements are
important in conjunction with frequency-domain spectral
studies \cite{Craig,Roukes1,Pressure_dep,Carr,Max}. While the
latter yield important mechanical properties, the former can
be useful in many applications, including optimization of the
excitation parameters, aiding the identification of sites
most effective for localized functionalization to enable
sensing, and in studies of dissipation effects such as
intrinsic and pressure-dependent damping \cite{Pressure_dep}.
Space-domain profiling is crucial in the presence of
significant nonlinearities where boundary conditions become
critical \cite{RonLif}. For instance, in typical capacitive
electrical drive configuration, the force between the
grounded substrate and a device fabricated by patterned
suspended poly-crystalline silicon (polySi) film (serving as
an electrode) is inherently nonlinear with the drive
amplitude and film stress
\cite{Craig,Roukes1,Pressure_dep,Carr,Max,RonLif}.

With interferometric reflection-mode optical transduction
\cite{Carr}, thermoacoustic effects can significantly modify
the effective device stiffness or induce autoparametric
optical drive \cite{Max}. We observe all these effects to be
significant in NEMS/MEMS devices defined on films with low
compressive residual stress under applied loads near to the
Euler critical value \cite{Buckled_exp}. At the critical
point, MEMS devices are most sensitive to changes induced by
stress variations in chemically-reactive coatings
\cite{SensorReview,Darren}. Of additional practical interest
are possible non-uniformities of mechanical and electrical
film properties across the wafer, originating from growth
processes and application of anisotropic etch, which directly
affect each circumferentially-clamped microresonator ($\mu$R)
\cite{Buckled_exp}. Fast space-domain visualization of
resonant VPs serves as a direct means to study the physics of
all these effects on the single device micro-scale during its
actuation.

VPs in MEMS are traditionally imaged optically with
vibrometric \cite{Pressure_dep}, interferometric
\cite{intMZ,intHeterodyne}, or stroboscopic \cite{Strobo,YL}
microscopy. Recently, spatiotemporal evaluation of resonant
VPs in high-frequency MEMS $\mu$Rs was demonstrated using
resonant realtime synchronous imaging (RSI) with a pulsed low
duty-cycle nanosecond laser \cite{YL}. The main feature of
stroboscopic imaging is a rapid production of time-resolved
interference pattern movies and static profiles, as well as
the fast evaluation of VPs, thus supplanting scanned
motorized probes that are expensive and inherently slow. This
technique is applicable with mechanical resonant frequencies
up to $f_{0}\simeq1$GHz. In this paper, we use RSI \cite{YL}
in averaging mode to rapidly characterize the VPs in bridge
$\mu$Rs close to critical stress as a function of the drive
amplitude and ambient pressure. The effects of high drive
nonlinearity and air damping on the resonator VPs are
directly monitored.

\section{Experimental method}

$\mu$Rs are fabricated by standard top-down micromachining
methods, where bridges are defined photolithographically on
compressively-stressed n+ doped polySi films, deposited by
low pressure chemical vapor deposition over a sacrificial
oxide layer and wet-etch released. Upon release of doubly
clamped bridges, residual stress is relieved through buckling
\cite{Buckled_exp,Darren}. The devices are driven
capacitively with the moving $\mu$R serving as an electrode
and the silicon substrate serving as a bottom ground
electrode. The inset in Fig. 1(a) shows the schematics of a
buckled $\mu$R cross section, as well as definitions we use,
and Fig. 1(b) shows SEM images of our released bridges.

\begin{figure}[t]
\includegraphics[width=8cm]{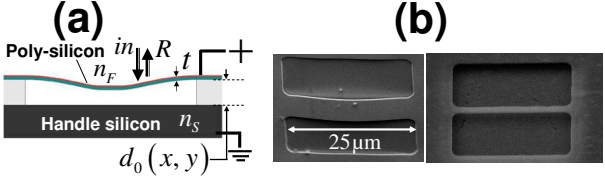}
\caption{(a) Schematic cross section of the devices studied
and definitions of optical quantities used in the analysis.
(b) SEM images of bridges of dimensions:
25$\times$6$\times$0.12 $\mu$m$^{3}$ (left, slightly
post-buckled), and 20$\times$1$\times$0.14 $\mu$m$^{3}$
(right, flat), both with $\sim$220 nm static elevations.}
\end{figure}

\begin{figure}[b]
\includegraphics[width=8cm]{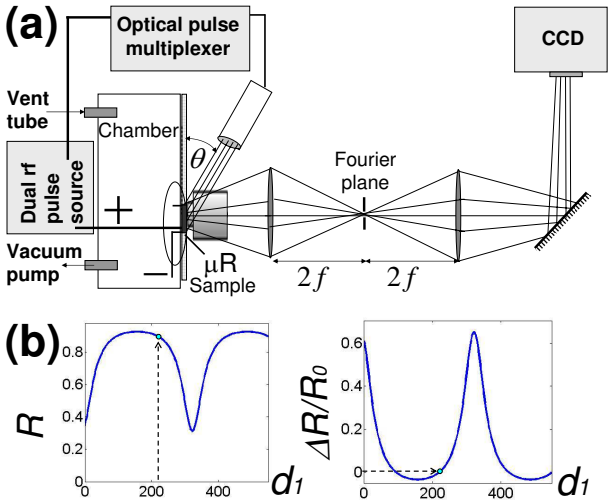}
\caption{(Color online) (a) Schematics of the experimental
setup. (b) Calibration curves for synchronous imaging
assuming a device with film thickness $t$=138 nm and static
midpoint elevation $d_{0}$=220 nm. Left: absolute reflection
coefficient $R$. Right: differential reflection $\Delta R /
R_{0}$. With negative values of $\Delta R$, the intensity
contrast in the image is negative.}
\end{figure}

In Fig. 2(a) a schematic of our RSI configuration is
illustrated. A dual channel pulse source feeds the $\mu$R and
optical imaging pulse source ($\lambda_{0}$=661.5 nm) in
synchrony. The collimated illumination at a glancing angle
$\theta\simeq40^{\circ}$ is reflected off the $\mu$R and
collected by an objective lens followed by a $4f$ lens pair,
the latter of which is used for spatial waveform filtering at
the Fourier plane with a phase mask \cite{Fourier}. The
outgoing light is finally imaged on a standard CCD camera.
Changes in the reflection with respect to the static image of
the $\mu$R, due to resonant motion, are monitored as a
function of the rf source frequency $f_{0}$, voltage and
phase. The pressure within the chamber is set with a vacuum
pump and venting tubes, and monitored via a Pirani gauge. In
order to calibrate the physical VPs from measured reflection
images, an interferometric analysis is carried out in the
out-of-plane direction (shown in Fig. 1(a)), as detailed
below. Application of a 50\% duty-cycle to the imaging pulses
(full synchronization with the capacitive drive), high
in-phase sensitivity to \emph{average} differential actuation
amplitudes is attained at the expense of lost temporal
resolution.

For calibration of the physical VPs from measured
reflectivity images, a Fabry-P\`{e}rot interferometer
multilayer analysis, as a function of the total elevation
\cite{Interference_book}, is performed using knowledge of the
static film elevation profile $d_{0}$, thickness $t$, and the
refractive indices of the film (n-doped polySi, $n$=3.916)
and substrate $n_{S}$ (single crystal Si, $n$=3.834). The
reflectance coefficients are calculated from the effective
reflectivity matrix, assuming nearly normal incidence:

\begin{eqnarray}
M_{total}=M_{2} \cdot M_{1}= \left(
                               \begin{array}{cc}
                                 \cos \delta_{2} & \frac{i}{n_{2}} \sin \delta_{2} \\
                                 in_{2} \sin \delta_{2} & \cos \delta_{2} \\
                               \end{array}
                             \right) \cdot \nonumber\\
                             \left(
                               \begin{array}{cc}
                                 \cos \delta_{1} & \frac{i}{n_{1}} \sin \delta_{1} \\
                                 in_{1} \sin \delta_{1} & \cos \delta_{1} \\
                               \end{array}
                             \right)
\label{eqs}
\end{eqnarray}

\begin{figure*}
\includegraphics[width=18cm]{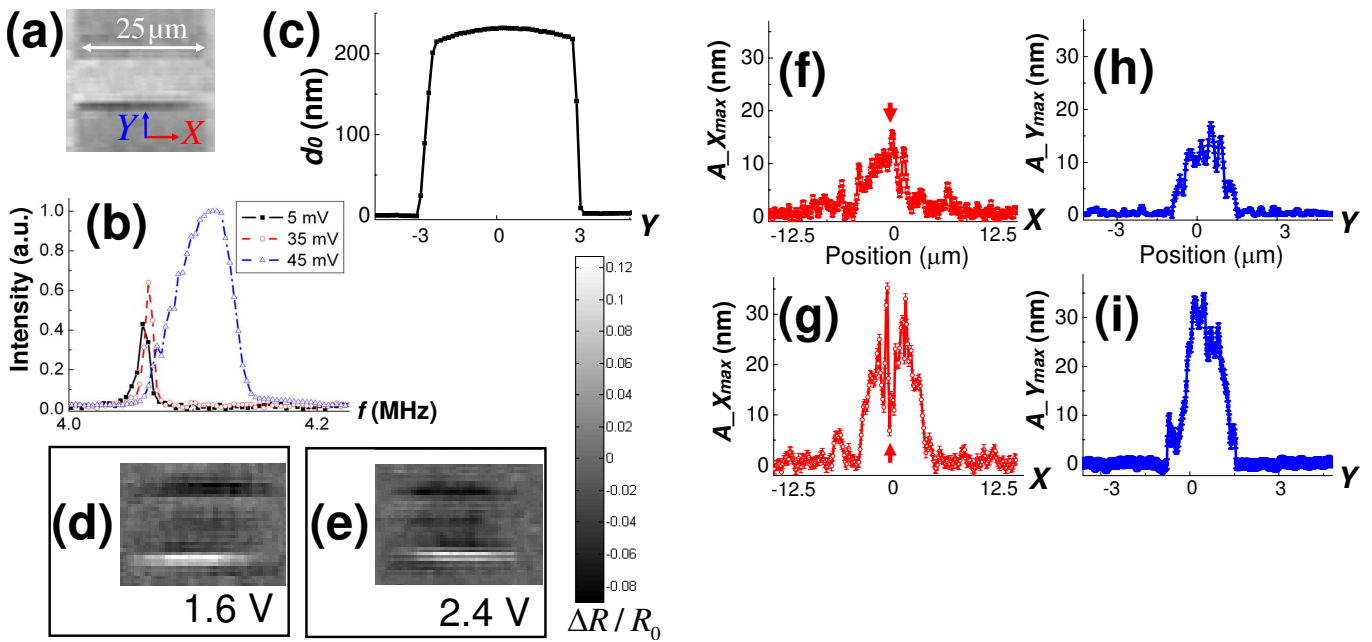}
\caption{\label{fig:epsart}(Color online) Drive amplitude
dependence of VPs in the fundamental mode of a critically
upward buckled resonator. (a) Reference image of the static
bridge. (b) Frequency domain spectra with low ac actuation
voltage and a constant 5 V dc bias. (c) Static height profile
of the bridge along the $Y$ direction taken from AFM
measurements (0 is defined as the height of the trench and
known film thickness of 140 nm is subtracted on the bridge).
(d),(e) Measured synchronous images at $f_{0}$=4.1 MHz with
different ac amplitudes; (f)-(i) corresponding VPs integrated
along $Y$ [in (f),(g), $X$-profiles], and along $X$ [in
(h),(i), $Y$-profiles]. Vertical arrows in (f),(g) indicate
diminished actuation signals with high excitation.}
\end{figure*}

\noindent where $\delta_{j}=k_{j}d_{j}$ is the effective
phase of layer $j$ and $k_{j}$ is the wave number. Denoting
$R_{0}(x,y)$ the reference image of the static reflection,
the contrast signal measured during actuation corresponds to:

\begin{equation}
R_{meas}(x,y) = \frac{R(x,y)-R_{0}(x,y)}{R_{0}(x,y)} \equiv
\frac{\Delta R(x,y)}{R_{0}}
\end{equation}

Under full synchronization of the sampling beam with the
drive frequency and phase, the observed average amplitude
$<A>$ at transverse position $(x,y)$ is then:

\begin{equation}
\langle A \rangle (x,y) = d_{1}(x,y) - d_{0}(x,y)
\end{equation}

With full sampling synchronization and a periodic unipolar
square wave excitation, the observed average amplitude
$\langle A \rangle$ at transverse position $(x,y)$ is $
\langle A \rangle=A_{max}/2$. The glancing angle of
illumination (see Fig. 1(a)) introduces an additional scaling
factor $\cos\theta$. The peak VP average amplitude, as
measured away from the static beam position, is then:

\begin{equation}
A_{max}(x,y)=\frac{2}{\cos\theta}[d_{1}(x,y)-d_{0}(x,y)]
\end{equation}

\noindent giving rise to a normalization factor of 2.83 in
our implementation, with $\theta=45^{\circ}$. A glancing
angle also introduces shadow effects at the resonator edges,
which we can easily eliminate with appropriate phase masks at
the Fourier plane (see Fig. 1(a)). We consider only
reflection variations at the positions of the $\mu$R itself
to constitute its real VPs. An example of the total and
differential reflectance curves as function of the total
elevations is shown in Fig. 2(b). Using Eqs. (1)-(4),
together with the calibration curve, the average maximum
amplitude profiles are estimated. Here we will concentrate on
characterizations of the fundamental (lowest) resonant mode.

\begin{figure*}
\includegraphics[width=16cm]{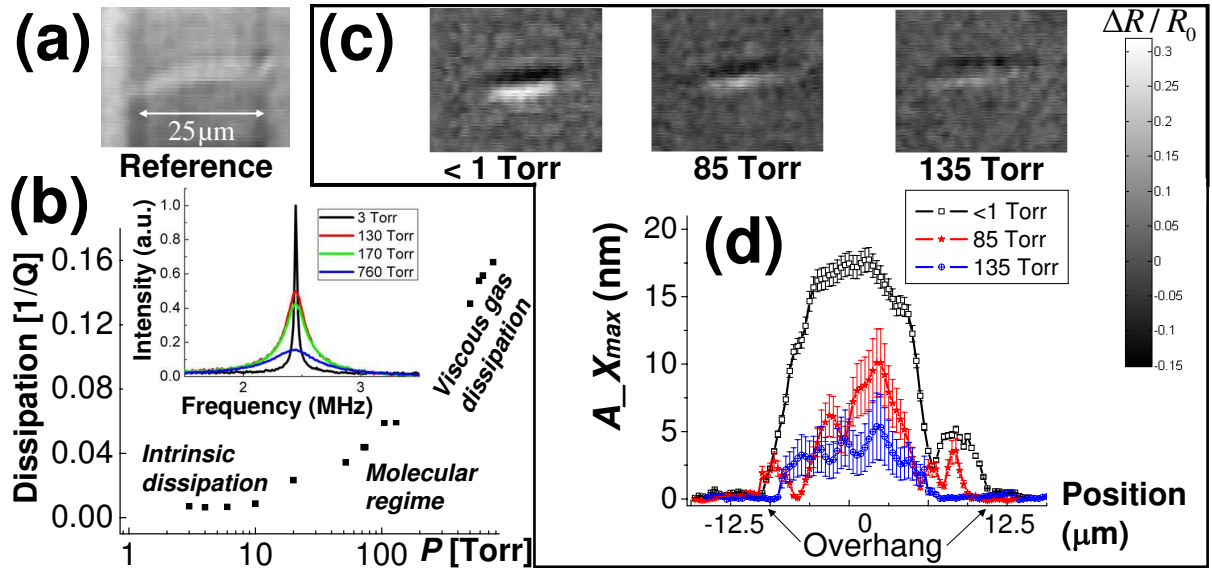}
\caption{\label{fig:epsart}(Color online) Pressure-dependent
studies of VPs in the fundamental mode of a slightly
post-buckled resonator. (a) Reference image of the static
bridge. (b) Frequency domain spectra (inset) and inverse
quality factors (dissipation) as a function of chamber
pressure, under continuous 315 mV ac drive and a 5 V dc bias.
(c) Measured interferometric images at $f_{0}$=2.4 MHz as a
function of the pressure, and (d) overlaid $Y$-integrated
$X$-profiles of vibration.}
\end{figure*}

\section{Results and Discussion}

In Fig. 3 we study a $\mu$R of dimensions
(25$\times$6$\times$0.14 $\mu$m$^{3}$) and a midpoint
elevation of 220 nm under low pressure settings (P$<$1 Torr).
The undriven $\mu$R is almost flat (see Fig. 3(c)) whereas
other slightly longer devices exhibit noticeable static
upward buckling, suggesting the existence of a compressive
force whose magnitude is close to critical load. Highly
buckled resonators have been found as hard to drive
electrostatically. Figure 3(a) shows a static optical image
of the unactuated device in its initial reference
configuration. In Fig. 3(b), the frequency response under
low-voltage actuation is shown. Even with drive amplitudes as
low as 45 mV and a dc bias of 5 V, we observe the formation
of Duffing nonlinearity \cite{RonLif} and significant
spectral broadening, with a sudden frequency detuning between
35 and 45 mV drive voltage. An AFM measurement of the static
bridge height profile, in the transverse ($Y$) direction, is
shown in Fig. 3(c), and the profile is uniform in the axial
($X$) direction, indicating a shell-like bridge profile. RSI
images with intermediate and high ac drive voltages, at a
frequency corresponding to the maximum resonant amplitude,
optimal phase and a dc bias of 5 V, are shown in Figs. 3(d)
and 3(e), respectively. Following the image analysis for
amplitude calibration, as detailed in the experimental
section, and integration along the beam width ($Y$), the peak
VP amplitude $X$-profiles are shown in Figs. 3(f) and 3(g),
respectively. The peak VP amplitude $Y$-profiles, integrated
and averaged along $X$, are also shown in Figs. 3(h) and
3(i). With intermediate drive amplitudes the VP shapes are as
shown in Fig. 3(f) and 3(h). With high drive amplitudes,
central regions on the beam appear to undergo diminished
displacement at the original frequency (Fig. 2(g)). However,
detuning of the imaging frequency in these cases to values
near multiples of the fundamental mechanical frequency and
the same phase settings show some tiny components of
vibration at these locations. We interpret this observation
as resulting from either nonlinear electromechanical
processes inducing transfer of energy to higher harmonics at
locations of high vibration amplitudes on the $\mu$R, or from
optical nonlinearity due to the measured response crossing
extreme reflection points. In any case, positions with
diminished signal, such as the one indicated by the vertical
arrows in Figs. 3(f),(g) would clearly not be beneficial to
employ in phase-locked-loop (PLL) sensing applications, using
this class of $\mu$Rs, at this wavelength. Along the
$Y$-profiles, slight localization of the motion at the
central region of the bridge is also observed with high
drives.

Figure 4 shows studies using a narrow microbridge of
dimensions (25$\times$1$\times$0.12 $\mu$m$^{3}$) and a
midpoint elevation of 660 nm ($t$=120 nm and $d_{0}$=660 nm)
under varying ambient hydrostatic pressure and constant
driving conditions of 1.2 V ac voltage and 5 V dc bias. This
bridge is slightly buckled in the upward direction, as
observed in the static optical reference image of Fig. 4(a).
Figure 4(b) shows the dissipation (inverse quality factor,
$Q^{-1}$) of the fundamental resonant mode as a function of
pressure, and corresponding spectra (inset). Different
pressure ranges correspond to well-known dominant dissipation
mechanisms\cite{Pressure_dep,Darren2}. The total quality
factor $Q$ is known to approximately scale according to
\cite{Darren2}:
\begin{equation}
1/Q = 1/Q_{int} + \alpha P
\end{equation}
\noindent Where $Q_{int}$ is the intrinsic (material) quality
factor, $\alpha$ is the coefficient of viscous damping and
$P$ is the pressure. In the data corresponding to Fig. 4(b),
a linear fit yields $Q_{int} = 154$ and $\alpha = 1.83 \times
10^{-4}$
 $[\textmd{Torr}^{-1}]$ in this $\mu$R.

In the current experiment we have succeeded in recording RSI
images of sufficient contrast only at pressures below the
viscous (gas-dominated) regime, namely, corresponding to the
intrinsic and molecular regimes in Fig. 4(b). It is estimated
that the most significant limiting factors are the low
spectral signal-to-noise bandwidth (S/N) at low quality
factors (below $Q\approx$ 20) combined with diminished
amplitudes of motion under external air damping. Figure 4(c)
shows RSI images of the $\mu$R as a function of increasing
pressure, with a transition from intrinsic to molecular
damping. Calibrated maximum amplitude $X$-profiles,
integrated across the beam width ($Y$), are shown in Fig.
4(d). Increased errors in the VP estimations result from
diminished available S/N, giving rise to less accurate
numerical fits. We consistently find that with increasing
pressure, the VPs in this $\mu$R become suppressed around the
regions close to the bridge overhang (see Fig. 4(d)). This
edge suppression effect is not observed in repeated
experiments under low pressure and drive conditions (0.3 V ac
voltage and 5 V dc bias), that yield available
signal-to-noise close to the detection limit, with extracted
vibration amplitudes comparable to the highest pressure case
shown here and with more pronounced motion near the overhang.

\section{Numerical Model}

The dynamics of a compressively stressed beam are described
by the equation \cite{Nayfeh,Slava}:

\begin{eqnarray*}
EI(\frac{\partial^{4}w}{\partial
x^{4}}-\frac{\partial^{4}w_{0}}{\partial
x^{4}})-[P-\frac{EA}{2L}\int^{L}_{0}((\frac{\partial
w}{dx})^{2}-(\frac{\partial w_{0}}{\partial x})^{2})dx]\times \nonumber\\
\frac{\partial^{2}w}{\partial x^{2}} +\rho d
\frac{\partial^{2}w}{\partial t^{2}} =
-\frac{\epsilon_{0}bV^{2}}{2(g_{0}+w)^{2}}\label{eqs}
\end{eqnarray*}

\begin{figure}[t]
\includegraphics[width=8cm]{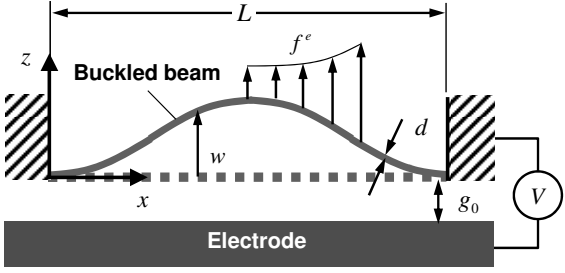}
\caption{Schematics of the numerical model.}
\end{figure}

\noindent where now $E$ is the Young's modulus of the beam
material, $I=b \times d^{3}/12$ is the moment of inertia of
the beam cross section, $A=b \times d$ is the sectional area,
$L$ is the beam length, $\rho$ is the density, and $b$ and
$d$ are the thickness and width of the beam, respectively. In
addition, $g_{0}$ is distance between the ends of the flat
side of the beam and the electrode (electrostatic gap),
$\epsilon_{0}$ is the vacuum permittivity and $V(t)$ is the
time-dependent actuation voltage. In accordance with the
definitions in Fig. 5, the elevation of the beam $w(x)$, as
well as the electrostatic force
$f^{e}(x,t)=-\epsilon_{0}bV^{2}/2(g_{0}+w)^{2}$, which is
calculated using the simplest parallel capacitor
approximation formula, are considered positive upwards.

Equation (5) has been reduced to the system of coupled
nonlinear ordinary differential equations by means of the
Galerkin decomposition with linear undamped eigenmodes of a
straight beam used as base functions. The equaations were
solved numerically using the ODE45-solver implemented in
Matlab. The details of the formulation and numerical approach
used for the analysis are found in \cite{Slava} (see also
\cite{Nayfeh}).

\section{Numerical Results}

Figures 6-8 show numerical solutions of Eq. (5). In all
cases, the actuation voltage contained both ac and dc bias
components, such that $V(t)=V_{dc}+V_{ac}\cos(\omega t)$.
Zero initial conditions, corresponding to the post-buckled
configuration of the beam in rest, were used. In all cases,
Young's modulus $E$=150 GPa and density $\rho$=2300
kg/m$^{3}$ corresponding to polySi were used.

\begin{figure}[b]
\includegraphics[width=8cm]{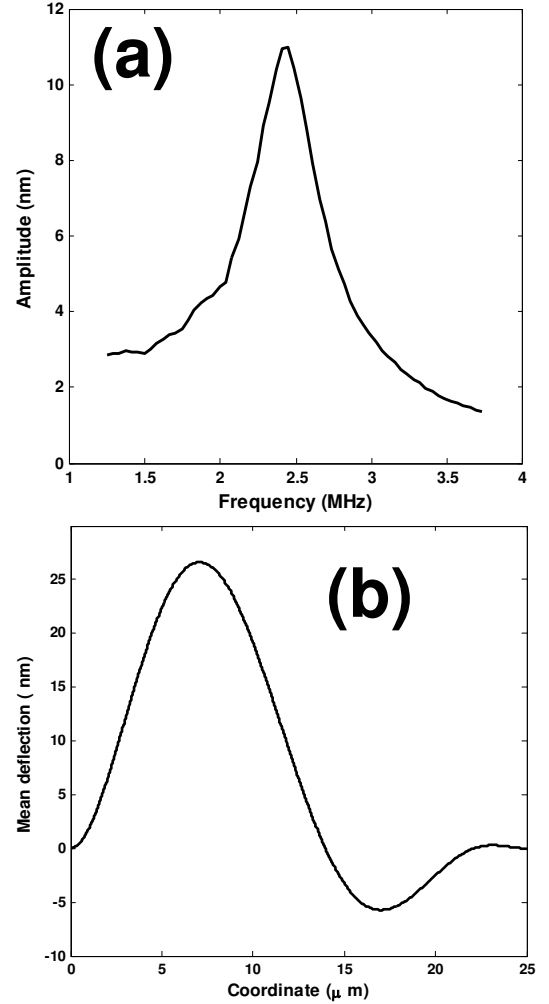}
\caption{(a) Numerical frequency domain resonant response.
Midpoint deflection of the beam is shown. (b) Deflection
profiles (difference between the actual and initial
elevations of the beam) averaged over a single period. In (b)
the operation frequency, $\omega$=2.45 MHz, is close to the
fundamental mode resonant frequency. In both simulations, the
dimensions of the beam are 25$\times$1$\times$0.12
$\mu$m$^{3}$; the electrostatic gap is $g_{0}$=660 nm; the
initial elevation of the midpoint above the beam's ends, due
to buckling, is 155 nm; input voltages are $V_{dc}=5$ V and
$V_{ac}=1.2$ V, and (a) $Q$=7; (b) $Q$=700.}
\end{figure}

Figure 6(a) shows the resonant response of the beam with
dimensions (25$\times$1$\times$0.12 $\mu$m$^{3}$) and
electrostatic gap $g_{0}$=660 nm. The axial force was chosen
such that the midpoint elevation of the beam above its ends
was 155 nm. The driving voltages were $V_{dc}=5$ V and
$V_{ac}=1.2$ V, and the quality factor was $Q$=7. It is
observed that the fundamental resonant frequency is 2.45 MHz,
which is close to the experimentally observed value (Fig.
4(b)). The corresponding resonant displacement profile,
averaged over a single period, is shown in Fig. 6(b). Small
initial imperfection of 0.05 in the initial buckled height,
corresponding to an excitation of the second anti-symmetric
mode, was introduced in order to allow non-symmetric mode
shapes of the beam. Calculations show that while the actual
beam profiles are dominated by the fundamental mode, the
resonant deflection profiles (i.e., the differences between
the initial buckled shapes and the actual, time dependent
shapes of the vibrating beam) could be more complex.

\begin{figure*}
\includegraphics[width=16cm]{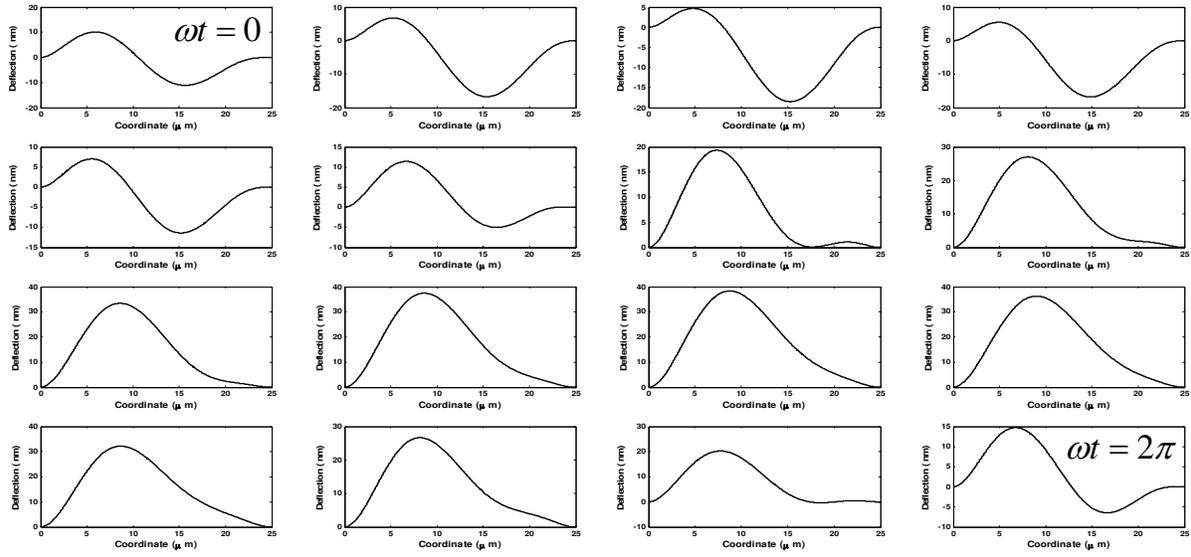}
\caption{\label{fig:epsart}Dynamic snapshots of the
deflection profiles (differences between actual and initial
elevations of the beam) corresponding to different time
sections within a single resonant cycle. The beam dimensions
are 25$\times$1$\times$0.12 $\mu$m$^{3}$; the electrostatic
gap is $g_{0}$=220 nm; the initial elevation of the midpoint
above the beam's ends, due to buckling, is 98 nm; input
voltages are $V_{dc}=1$ V and $V_{ac}=350$ mV, and $Q$=1000.
The operation frequency is $\omega$=1.495 MHz. Nine
(symmetric and skew-symmetric) base functions are preserved
in the reduced-order model.}
\end{figure*}

A decrease in the initial separation between the beam and the
electrode results in an increased contribution of higher
modes in the resonant VPs. Figure 7 shows time-resolved
snapshots of the VPs (relative displacements from
equilibrium) with the same dimensions as in Fig. 6, but with
$g_{0}$=220 nm; Figure 8 shows the same vibration profile as
averaged over a single period. Small initial imperfection of
0.05 in the initial buckled height, corresponding to a
contribution of the second buckled mode, was again used as an
initial condition. Complex displacement profiles are clearly
observed in this case as well.

\begin{figure}[b]
\includegraphics[width=8cm]{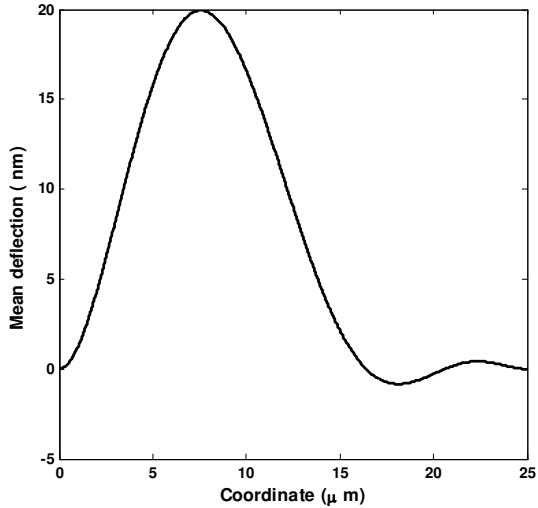}
\caption{Deflection profiles (difference between the actual
and initial beam elevations) averaged over a single period,
corresponding to the results in Fig. 7 (same parameters).}
\end{figure}

\section{Conclusion}

Synchronous imaging has been demonstrated as a robust method
for direct and rapid observations of gradual changes in
resonant vibration profiles of electromechanical
microresonators under varying conditions of drive and ambient
pressure. Synchronous imaging can serve as a useful tool for
studying fundamental processes in resonant MEMS/NEMS, as well
as for identification of favorable device positions most
suitable for sensitive phase-locked applications, such as
sensors, filters and switches. Numerical analysis of
electrostatically driven post-buckled microbridges supports
the richness of the complex resonant vibrations that are
possible in these micro-electromechanical systems.

\section{Acknowledgments}

This research was funded by the National Science Foundation
(grants DMR-0908634 and DMR-0520404) and Analog Devices.
Fabrication was performed at the Cornell Nanoscale science
and technology Facility.


\begin{references}

\bibitem[Craig(2000)]{Craig} H. G. Craighead, Science \textbf{290}, 5496 (2000).
\bibitem[Roukes1(2005)]{Roukes1} K. L. Ekinci and M. L. Roukes, Rev. Sci. Instrum. \textbf{76}, 061101 (2005).
\bibitem[Pressure_dep(2011)]{Pressure_dep} R. C. Tung RC, J. W. Lee, H. Sumali, and A. Raman, J. Micromech. Microeng. \textbf{21}, 025003
(2011); R. A. Bidkar, R. C. Tung, A. A. Alexeenko, H. Sumali,
and A. Raman, Appl. Phys. Lett. \textbf{94}, 163117 (2009);
H. Sumali, J. Micromech. Microeng. \textbf{17}, 2231 (2007).

\bibitem[Carr(1999)]{Carr} D. W. Carr, S. Evoy, L. Sekaric, H. G. Craighead, and J. M.
Parpia, Appl. Phys. Lett. \textbf{75}, 920 (1999).
\bibitem[Max(2000)]{Max} M. Zalalutdinov, A. Zehnder, A. Olkhovets, S. Turner, L. Sekaric, B. Ilic, D. Czaplewski, J. M. Parpia, and H. G. Craighead, Appl. Phys. Lett. \textbf{79}, 695
(2001); B. Ilic, S. Krylov, K. Aubin, R. Reichenbach, and H.
G. Craighead, Appl. Phys. Lett. \textbf{86}, 193114 (2005).
\bibitem[RonLif(2008)]{RonLif} R. Lifshitz and M. C. Cross, \emph{Review of Nonlinear Dynamics and Complexity} (Wiley, Meinheim, 2008), Vol. I, pp. 1-52.
\bibitem[Buckled_exp(1999)]{Buckled_exp} W. Fang, C.-H. Lee, and H.-H. Hu, J. Micromech. Microeng. \textbf{9}, 236 (1999).
\bibitem[SensorReview(2011)]{SensorReview} A. Boisen, S. Dohn, S. S. Keller, S. Schmid, and M. Tenje, Rep. Prog. Phys. \textbf{74}, 036101 (2011).
\bibitem[Darren(2010)]{Darren} D. R. Southworth, L. M. Bellan, Y. Linzon, H. G. Craighead, and J. M. Parpia, Appl. Phys. Lett. \textbf{96}, 163503 (2010).
\bibitem[intMZ(2001)]{intMZ} G. G. Fattinger and P. T. Tikka, Appl. Phys. Lett. \textbf{79}, 290 (2001).
\bibitem[intHeterodyne(2008)]{intHeterodyne} K. Kokkonen and M. Kaivola, Appl. Phys. Lett. {\bf 92}, 063502 (2008).
\bibitem[Strobo(2002)]{Strobo} C. Rembe and R. S. Muller, J. Microelectromech S. \textbf{11}, 479 (2002).
\bibitem[YL(2010)]{YL} Y. Linzon, S. Krylov, B. Ilic, D. R. Southworth, R. A. Barton, B. R. Cipriany, J. D. Cross, J. M. Parpia, and H. G. Craighead, Opt. Lett. \textbf{15}, 2654 (2010).
\bibitem[Darren2(2009)]{Darren2} D. R. Southworth, H. G. Craighead, and J. M. Parpia, Appl. Phys. Lett. \textbf{94}, 213506 (2009).
\bibitem[Fourier(1978)]{Fourier} J. D. Gaskill, \emph{Linear Systems, Fourier Transforms, and Optics} (Wiley, New York, 1978).
\bibitem[Interference_book(1995)]{Interference_book} M. Bass, \emph{Handbook of Optics}, 2nd ed. (McGraw-Hill, San Francisco, 1995), Vol. I, pp. 42.10-42.14.
\bibitem[Slava(2011)]{Slava} S. Krylov, B. R. Ilic, and S. Lulinsky, Nonlinear Dyn. \textbf{66}, 403 (2011).
\bibitem[Nayfeh(2004)]{Nayfeh} S. A. Emam and A. H. Nayfeh, Nonlinear Dyn. \textbf{35}, 1 (2004).

\end{references}
\end{document}